\documentclass[conference]{IEEEtran}
\IEEEoverridecommandlockouts
\usepackage{cite}
\usepackage{amsmath,amssymb,amsfonts}
\usepackage{algorithm}
\usepackage{algpseudocode}
\usepackage{graphicx}
\usepackage{textcomp}
\usepackage{xcolor}
\usepackage{booktabs}
\usepackage{url}
\usepackage{balance}

\def\BibTeX{{\rm B\kern-.05em{\sc i\kern-.025em b}\kern-.08em
    T\kern-.1667em\lower.7ex\hbox{E}\kern-.125emX}}
\begin{document}
\title{Adaptive and Proximal Federated Optimization\\ for Transferable Network Intrusion Detection}
\author{
\IEEEauthorblockN{Rezwan Islam Salvi, MD Rafin Jawad Hafiz, Abeer Mahmood Kalam, and Md Moniruzzaman}
\IEEEauthorblockA{\textit{Department of Computer Science and Engineering}\\
\textit{Islamic University of Technology}\\
Dhaka, Bangladesh\\
\{rezwanislam, rafinjawad, abeermahmood, milton\}@iut-dhaka.edu}
}
\maketitle
\begin{abstract}
Network Intrusion Detection Systems (NIDS) based on deep learning often fail to detect attack classes that are absent from training data, limiting their robustness against zero-day threats. This property called transferability across attack classes, remains underexplored especially in federated learning (FL) settings where data heterogeneity and client drift pose additional challenges. Prior work (TabFIDS) showed that FedAvg combined with bootstrapping and temporal averaging improves transferability for cross-device NIDS. In this paper, we explore whether other federated optimizers can do better. We compare two optimizers inspired by FedAdam and FedProx against the FedAvg baseline on the CIC-IDS 2017 dataset using an 11-client federated setup, sweep 10 hyperparameter configurations, and contrast two data-balancing strategies: bootstrapping and SMOTE. Using the count of client pairs whose transferred-model accuracy exceeds 70\,\% as our primary metric (following TabFIDS), the proximal optimizer with a strong regularization term yields 48 transferable pairs versus 42 for FedAvg, while the adaptive optimizer with an aggressive server learning rate yields 46 pairs. SMOTE improves per-pair accuracy over bootstrapping but reduces breadth, revealing a depth-vs-breadth trade-off: high accuracy on a few client pairs versus moderate accuracy across most. These results indicate that the choice of federated optimizer, the rule by which client updates are regularized and aggregated into the global model, is very important for federated NIDS.
\end{abstract}

\section{Introduction}
Network Intrusion Detection Systems (NIDS) are essential for securing modern networks against an evolving landscape of cyber threats~\cite{sharafaldin2018toward}. Deep learning has substantially improved NIDS performance on known attacks by automatically learning intricate, non-linear traffic patterns, eliminating the heavy manual feature engineering required by classical methods such as Naive-Bayes classifiers, Random Forests, and Support Vector Machines~\cite{ghosh2023study}. Yet trained detectors still struggle to generalize to attack classes that were absent from training data. This generalization property---\emph{transferability}---is critical for detecting zero-day threats without frequent retraining. A model that transfers well across attack classes also reduces computational cost, since useful detectors can be learned from a fragment of the threat space rather than a large corpus.

A second challenge in deploying NIDS at scale is data privacy. Federated learning (FL)~\cite{mcmahan2017communication} addresses this by training a global model across distributed clients without sharing raw data, keeping sensitive records on-device. However, FL's effectiveness is sensitive to non-IID data distributions and client drift~\cite{kairouz2019advances}, both of which are pronounced in NIDS, where attack distributions vary sharply across deployment sites. Recent work has shown that an FL-based NIDS: TabFIDS~\cite{ghosh2024improving}, achieves better transferability than centralized baselines by combining FedAvg with bootstrapping and temporal averaging on the CIC-IDS 2017 dataset, reporting 31 transferable pairs against 17 for the centralized model and 2 for baseline federated training.

However, FedAvg itself is known to converge slowly and unstably under heterogeneity. Two well-established alternatives: FedAdam~\cite{reddi2020adaptive} (adaptive server-side optimization) and FedProx~\cite{li2020federated} (client-side proximal regularization)---have been shown to outperform FedAvg in non-IID settings on standard benchmarks, and FedProx has been successfully applied to anomaly-based NIDS in Fed-ANIDS~\cite{idrissi2022fedanids}. Whether these advances translate into improved transferability across attack classes, a setting fundamentally different from same-class evaluation, has not been studied systematically. Our work builds directly on the work of Ghosh et al.~\cite{ghosh2023analysis,ghosh2023study,ghosh2024improving}: we retain their CIC-IDS 2017 cross-device protocol and transferability metric, but replace the fixed FedAvg aggregation rule with adaptive and proximal optimizers inspired by FedAdam and FedProx, treating the choice of federated optimizer, rather than the preprocessing pipeline alone, as the primary design variable.

\textbf{Contributions.} This paper makes three contributions:
\begin{itemize}
    \item We replace TabFIDS's FedAvg with adaptive and proximal optimizers inspired by FedAdam and FedProx and conduct a hyperparameter sweep across 10 configurations, evaluating cross-class transferability on CIC-IDS 2017 under an identical 11-client protocol.
    \item We compare bootstrapping against SMOTE for class balancing in federated training, and show that bootstrapping dominates SMOTE.
    \item We characterize a depth-vs-breadth trade-off and identify which choices favour each goal.
\end{itemize}

\section{Related Work}
\textbf{The transferability line we build on.} Our study is a direct continuation of a line of work by Ghosh, Jameel, and El~Gamal on transferability in NIDS. Their initial analysis~\cite{ghosh2023analysis} framed cross-class transferability under distributed deep learning, and the expanded centralized study~\cite{ghosh2023study} trained a model on one attack class and tested it on another using CIC-IDS 2017. They reported both symmetric and asymmetric relationships among Denial-of-Service (DoS) families, e.g., training on DoS GoldenEye enabled detection of DoS Hulk, DDoS, Slowhttptest, and Slowloris through shared mechanisms such as HTTP flooding, and used Recursive Feature Elimination to show that highly transferable pairs share dominant features with one another. Transferability was largely confined to DoS attacks, with little success for Botnet, Infiltration, or Web-Attack classes. The follow-up TabFIDS~\cite{ghosh2024improving} moved this to a federated setup with 11 clients across 110 ordered attack pairs: temporal averaging alone yielded 14 transferable pairs, and combining it with bootstrapping reached 31 pairs, outperforming centralized (17) and baseline federated (2) configurations, while attaining 95.36\,\% average accuracy for same-class testing. Crucially, since each client trains on only one attack class, the only place where knowledge of different attacks can mix is the server, when it combines the clients' model updates, so this aggregation step is what enables cross-class transfer. TabFIDS, however, always combined updates with plain FedAvg and never tested whether a different federated optimizer could share this knowledge more effectively, which is the gap our study addresses.

\textbf{Most recent extensions by the same authors.} Two recent works extend this line. TabFIDSv2~\cite{jameel2025developing} introduces a Block-Based Smart Aggregation (BBSA) algorithm that makes performance-driven, layer-wise decisions within FedAvg aggregation, together with a data-driven feature-elimination step, and demonstrates transferability across backbones and datasets. FetFIDS~\cite{ghosh2025fetfids} instead replaces the backbone with a transformer that uses feature embedding for edge federated deployment. Both modify the model or the layer-wise aggregation policy; in contrast, we keep the TabFIDS preprocessing and CNN backbone fixed and change the \emph{federated optimizer} itself. Our contribution is therefore complementary.

\textbf{Federated optimization under heterogeneity.} FedAvg~\cite{mcmahan2017communication} underpins most FL-IDS pipelines but degrades under non-IID data and client drift~\cite{kairouz2019advances}, a recurring theme in FL-IDS surveys~\cite{lavaur2022evolution,khraisat2024survey}. Reddi et al.~\cite{reddi2020adaptive} proposed the FedOpt family (FedAdagrad, FedYogi, FedAdam), embedding adaptive moment estimation in the server update and consistently surpassing FedAvg, FedAvgM, and SCAFFOLD on heterogeneous benchmarks. FedProx~\cite{li2020federated} instead adds a $\mu$-weighted proximal term that penalizes local drift. Idrissi et al.~\cite{idrissi2022fedanids} applied FedProx in Fed-ANIDS for anomaly-based NIDS, reporting simple-autoencoder F1-scores of 99.44\,\% (USTC-TFC2016), 92.73\,\% (CIC-IDS 2017), and 86.41\,\% (CSE-CIC-IDS 2018), with FedProx generally outperforming FedAvg, but for anomaly detection, and not cross-class transferability.

\textbf{Class balancing.} Class imbalance is acute in CIC-IDS 2017 (roughly 80\,\% benign). Bootstrapping and SMOTE~\cite{chawla2002smote} are the two dominant remedies; Ghosh et al.~\cite{ghosh2023study} found bootstrapping superior to SMOTE for transferability in centralized training, an observation we revisit in the federated setup.

To our knowledge, no prior work including the most recent extensions above, has explored adaptive and proximal federated optimizers in the specific setting of attack-class transferability for federated NIDS.

\section{Methodology}

\subsection{Problem Formulation}
We adopt the standard federated objective of minimizing a weighted average of client losses,
\begin{equation}
\min_{x\in\mathbb{R}^d} \; f(x) = \frac{1}{m}\sum_{i=1}^{m} F_i(x),
\end{equation}
where $F_i(x)$ is the local loss of client $i$, whose data distribution may differ arbitrarily from the others. FedOpt generalizes FedAvg by allowing a client optimizer Stochastic Gradient Descent (SGD) and a separate server optimizer that combines the client updates.

\subsection{Federated Optimizers}
\textbf{FedAvg} (baseline) aggregates clients' parameters via a sample-weighted average each round, with no server-side adaptive rescaling or explicit regularization for heterogeneity. Our two optimizers adapt the ideas of FedAdam and FedProx to this setting; we make the implementation choices explicit below.

\textbf{Adaptive aggregation (FedAdam-inspired).} Following FedAdam~\cite{reddi2020adaptive}, the server treats the aggregated client model as a pseudo-gradient and applies an Adam-style update (Algorithm~\ref{alg:fedadam}). Our implementation differs from canonical FedAdam in two respects: the pseudo-gradient is the sample-weighted average of client weights rather than of client, and the Adam moment estimates are re-initialized each round rather than accumulated across rounds. As a consequence, the per-round update reduces to a learning-rate-scaled adaptive step ($\eta$).

\textbf{Proximal optimization (FedProx-inspired).} Following FedProx~\cite{li2020federated}, each client augments its local objective with the proximal term ($\mu$), penalizing drift from the global model (Algorithm~\ref{alg:fedprox}); our aggregation additionally damps each client's contribution toward a reference model by $\mu$. Larger $\mu$ therefore produces a more conservative global model less affected by any single client's update.

\begin{algorithm}[t]
\caption{Adaptive server aggregation (FedAdam-inspired)}
\label{alg:fedadam}
\begin{algorithmic}[1]
\Require server learning rate $\eta$; stabilizer $\epsilon$;
         number of rounds $T$; client sample counts $\{n_i\}$
\Ensure  trained global model $x$
\State initialize the global model $x$
\For{$t = 0,1,\dots,T-1$} \Comment{one communication round}
  \State \textbf{Broadcast:} send current global model $x$ to every client
  \For{each client $i$ \textbf{in parallel}}
    \State $x_i \leftarrow$ result of training $x$ locally on client $i$'s data (SGD)
  \EndFor
  \State \textbf{Aggregate (pseudo-gradient):}
  \State \quad $g_t \leftarrow \dfrac{\sum_i n_i\, x_i}{\sum_i n_i}$
         \Comment{sample-weighted average of client \emph{weights}}
  \State \textbf{Server update (single adaptive step):}
  \State \quad $x \leftarrow x - \eta\,\dfrac{g_t}{\sqrt{g_t^{2}}+\epsilon}$
         \Comment{one Adam-style step}
\EndFor
\State \Return $x$
\State \textbf{Note:} a fresh Adam step is taken each round from zero moment estimates, so the
       update reduces to $x \leftarrow x - \eta\,\mathrm{sign}(g_t)$; the server LR
       $\eta$ is the only effective lever.
\end{algorithmic}
\end{algorithm}

\begin{algorithm}[t]
\caption{Proximal regularization (FedProx-inspired)}
\label{alg:fedprox}
\begin{algorithmic}[1]
\Require local learning rate $\eta_l$; proximal weight $\mu$; local steps $K$;
         number of rounds $T$; client sample counts $\{n_i\}$
\Ensure  trained global model $x$
\State initialize the global model $x$
\For{$t = 0,1,\dots,T-1$} \Comment{one communication round}
  \State \textbf{Broadcast:} $x_t \leftarrow x$; send $x_t$ to every client
  \For{each client $i$ \textbf{in parallel}}
    \State $x_i \leftarrow x_t$ \Comment{start local training from global model}
    \For{$k = 0,1,\dots,K-1$} \Comment{local SGD steps}
      \State $x_i \leftarrow x_i - \eta_l\big(\underbrace{g^{t}_{i,k}}_{\text{data gradient}}
             + \underbrace{\mu\,(x_i - x_t)}_{\text{proximal pull to global}}\big)$
    \EndFor
  \EndFor
  \State \textbf{Aggregate ($\mu$-damped):}
  \State \quad $x \leftarrow \dfrac{\sum_i n_i\big((1-\mu)\,x_i + \mu\,x_t\big)}{\sum_i n_i}$
         \Comment{weighted average, pulled toward $x_t$ by $\mu$}
\EndFor
\State \Return $x$
\State \textbf{Note:} the proximal weight $\mu$ acts twice, once in each client's local update
       via the term $\mu(x_i - x_t)$, and again as a damping during aggregation.
\end{algorithmic}
\end{algorithm}

\subsection{Why adaptive and proximal optimization}
Adaptive server optimization and proximal regularization are two remedies for FedAvg's instability under heterogeneity~\cite{reddi2020adaptive,li2020federated,idrissi2022fedanids}: one rescales the server step using gradient statistics, the other constrains local drift. Since each client here observes a single attack class which is an extreme non-IID setting, both are natural levers to test for transferability, and we study how their key parameters affect cross-class transfer.

\subsection{Data Balancing}
CIC-IDS 2017 is severely imbalanced (roughly 80\,\% benign), so balancing is essential. We compare two strategies. \textbf{Bootstrapping} resamples real attack data with Gaussian perturbations ($\sigma{=}0.01$) until matched with benign volume, following the protocol used in TabFIDS. \textbf{SMOTE}~\cite{chawla2002smote} generates synthetic minority samples by interpolation in feature space, which can sharpen minority-class decision boundaries without simply duplicating examples. Prior work~\cite{ghosh2023study} reported SMOTE underperforming in centralized training, but its behaviour in federated training where local class distributions are more skewed and shared structure across clients matters, has not been characterized for transferability.

\subsection{Hyperparameter Variants}
For the \textbf{adaptive optimizer} we sweep the server learning rate $\eta \in \{0.001, 0.005, 0.01, 0.02, 0.05\}$, yielding 5 configurations (v1--v5; Table~\ref{tab:fedadam}). v5 uses the most aggressive rate $\eta{=}0.05$. For the \textbf{proximal optimizer} we sweep the proximal weight $\mu \in \{0.001, 0.01, 0.1, 0.2, 0.5\}$ (Table~\ref{tab:fedprox}), with v4 corresponding to $\mu{=}0.5$.

\begin{table}[t]
\caption{Adaptive (FedAdam-inspired) variants: server \\ learning rate $\eta$ per variant; $\epsilon{=}10^{-8}$.}
\label{tab:fedadam}
\centering
\footnotesize
\begin{tabular}{@{}lcccccccccc@{}}
\toprule
\textbf{Variant} & v1 & v2 & v3 & v4 & v5 \\
\midrule
$\boldsymbol{\eta}$ & 0.001 & 0.005 & 0.01 & 0.02 & 0.05 \\
\bottomrule
\end{tabular}
\end{table}

\begin{table}[t]
\caption{Proximal (FedProx-inspired) weight variants.}
\label{tab:fedprox}
\centering
\small
\begin{tabular}{@{}lccccc@{}}
\toprule
\textbf{Variant} & v1 & v2 & v3 & v4 & v5 \\
\midrule
$\boldsymbol{\mu}$ & 0.01 & 0.1 & 0.2 & 0.5 & 0.001 \\
\bottomrule
\end{tabular}
\end{table}

\section{Experimental Setup}
We follow the publicly released implementation of the base TabFIDS work~\cite{ghosh2024improving}, available at \url{https://github.com/ghosh64/transferability}, and modify it for our needs: we replace FedAvg aggregation with our adaptive and proximal optimizers, add a SMOTE data balancing path alongside the original bootstrapping, and introduce the hyperparameter sweeps and 20\,\%/100\,\% data usage below.

\textbf{Dataset.} We use CIC-IDS 2017~\cite{sharafaldin2018toward}, a widely-used benchmark from the Canadian Institute for Cybersecurity capturing five days of realistic traffic, with over 2.8 million samples across 78 features and attack classes including DDoS, DoS variants, Brute Force, PortScan, and Web Attacks. Following~\cite{ghosh2024improving}, we drop classes with fewer than 100 samples and retain 11 attack classes.

\textbf{Preprocessing.} The eight CSV files are consolidated into a single frame. Infinite values are replaced with NaN and then imputed with zero; the 78 traffic features are normalized to $[0,1]$ with min--max scaling, and labels are encoded numerically. A custom dataset class then distributes data across clients, applying bootstrapping or SMOTE to balance benign and attack samples.

\textbf{Federated setup.} We simulate 11 clients (one per attack class) plus 1 central server, training for 20 communication rounds using NVIDIA RTX 3090 GPU, where each round is one cycle of the server broadcasting the global model, clients training locally on their own data, and the server aggregating the returned updates. Each client trains on its assigned attack class plus a partition of benign traffic, with no overlap of benign data across clients, matching the cross-device protocol of~\cite{ghosh2024improving}. We evaluate on both a 20\,\% stratified subset (for hyperparameter exploration) and the full 100\,\% dataset (for final evaluation).

\textbf{Transferability metric and accuracy calculation.} Following the transferability criterion introduced in TabFIDS~\cite{ghosh2024improving}, an ordered $(\text{source}, \text{target})$ pair is \emph{transferable} if a model trained on the source client achieves accuracy $>70\,\%$ on the target client's test data. As in~\cite{ghosh2024improving}, this accuracy is class-balanced: $\mathrm{Acc}=\tfrac{1}{2}\big(\tfrac{tn}{tn+fp}+\tfrac{tp}{tp+fn}\big)$, the average of both attack and benign detection rates measured for each pair. It measures how many actual attacks are caught ($\tfrac{tp}{tp+fn}$) and how much benign traffic is correctly caught as well ($\tfrac{tn}{tn+fp}$). Averaging the two gives the attack and benign classes equal weight, so a model cannot score well simply by labelling the dominant benign traffic of CIC-IDS 2017 correctly. We report the number of transferable pairs out of 110 ($11{\times}10$) and the average of the accuracy across these transferable pairs.

\section{Results}
\begin{figure}[t]
\centering
\includegraphics[width=0.99\columnwidth]{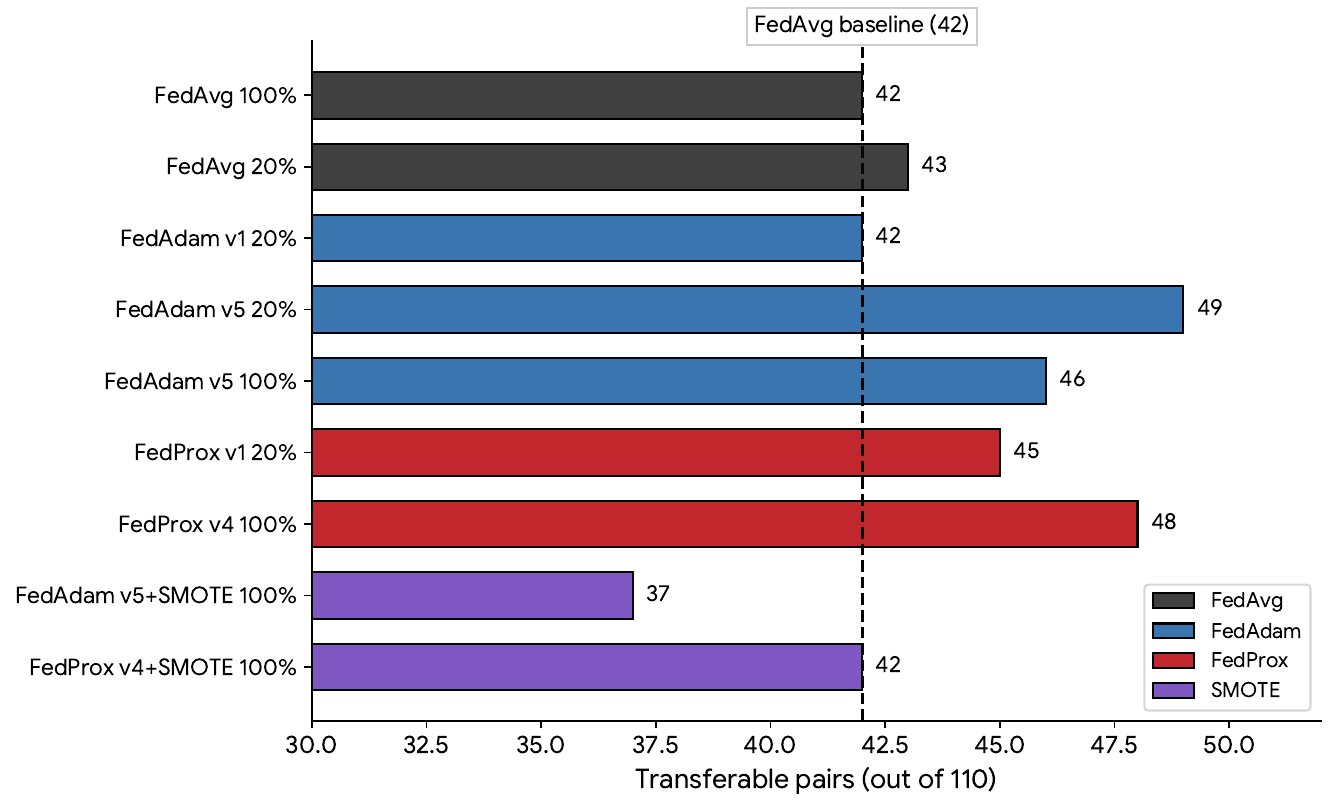}
\caption{Transferable pairs for the most significant configurations. We  label the FedAdam-inspired variants just `FedAdam' and the FedProx-inspired variants just `FedProx' for simplicity. The dashed line marks the FedAvg baseline (42). Both FedProx (v4) and FedAdam (v5) exceed it, while SMOTE variants fall below on breadth.}
\label{fig:bar}
\end{figure}

\begin{figure}[t]
\centering
\includegraphics[width=0.99\columnwidth]{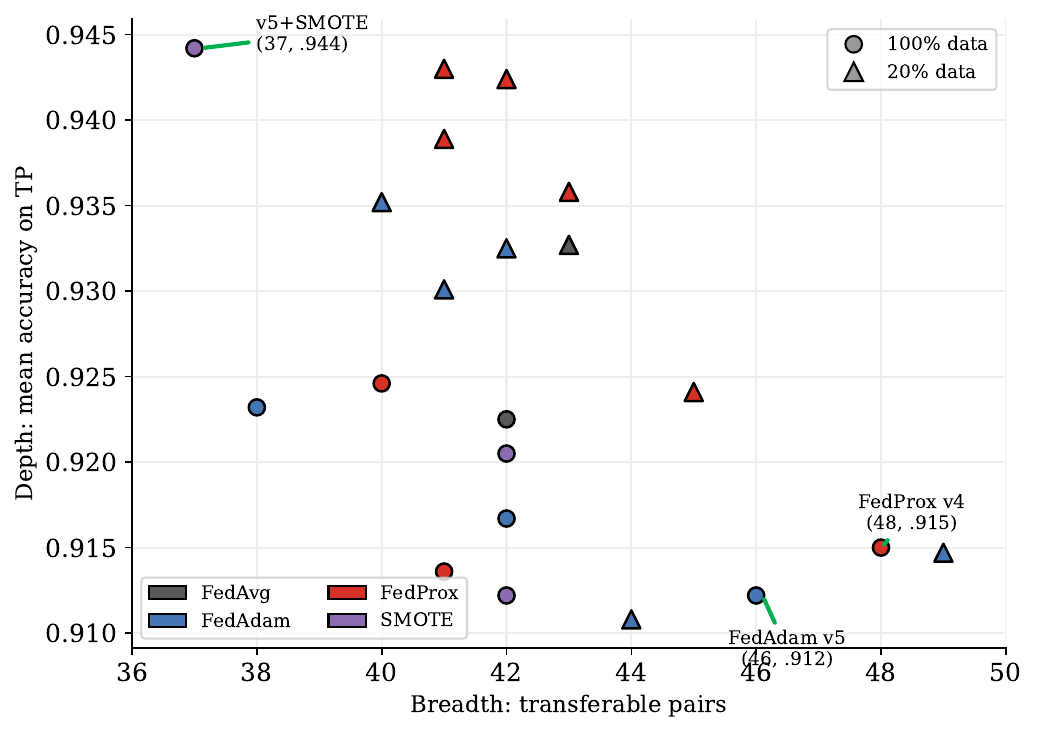}
\caption{Depth-vs-breadth view across all 21 runs. SMOTE configurations sit high (accuracy) but left (few pairs); FedProx v4 and FedAdam v5 push furthest right with bootstrapping (breadth).}
\label{fig:tradeoff}
\end{figure}

Table~\ref{tab:fullresults} reports all 21 runs; Fig.~\ref{fig:bar} highlights the most significant configurations and Fig.~\ref{fig:tradeoff} visualizes the depth-vs-breadth structure. We \textbf{label} the FedAdam-inspired variants just `FedAdam'
 and the FedProx-inspired variants just `FedProx' for simplicity. The FedAvg baseline reaches 42 transferable pairs at 0.923 average accuracy on the full dataset.

\textbf{FedProx} wins on breadth: v4 ($\mu{=}0.5$) achieves the highest full-dataset count with \textbf{48 transferable pairs}, a 14.3\,\% increase over FedAvg, while maintaining 0.915 average accuracy. \textbf{FedAdam} is competitive: v5 (server LR $0.05$) reaches 46 transferable pairs on the full dataset and 49 on the 20\,\% subset which is the highest count observed in any configuration.

\textbf{SMOTE} produces a clear trade-off. FedAdam v5 with SMOTE achieves the highest mean accuracy on transferable pairs (\textbf{0.944}) but only results in 37 transferable pairs---worse breadth than the same optimizer without SMOTE (46 pairs). The same pattern appears for FedAvg and FedProx with SMOTE: per-pair accuracy is preserved or slightly improved, but cross-client transferability is reduced. Bootstrapping consistently produces broader transferability than SMOTE, confirming the observation of~\cite{ghosh2023study} holds in our federated setup.

\begin{table*}[t]
\caption{Complete results across all configurations on CIC-IDS 2017. We  label the FedAdam-inspired variants just `FedAdam' and the FedProx-inspired variants just `FedProx' for simplicity. \#TP = transferable pairs (out of 110); Acc = mean accuracy on transferable pairs. Bold marks the highest \#TP overall (FedAdam v5, 49) and the best full-dataset \#TP (FedAdam v5, 46; FedProx v4, 48), plus the highest mean accuracy on transferable pairs (0.944). All are with bootstrapping except the last 3.}
\label{tab:fullresults}
\centering
\small
\begin{tabular}{@{}lcccc|lcccc@{}}
\toprule
\textbf{Method} & \textbf{Data} & \textbf{Runtime (hr)} & \textbf{\#TP} & \textbf{Acc} &
\textbf{Method} & \textbf{Data} & \textbf{Runtime (hr)} & \textbf{\#TP} & \textbf{Acc} \\
\midrule
FedAvg           & 100\,\% & 6.016 & 42 & 0.9225 & FedProx v1         & 100\,\% & 6.250  & 41 & 0.9136 \\
FedAvg           & 20\,\%  & 2.183 & 43 & 0.9327 & FedProx v2         & 20\,\%  & 2.900  & 42 & 0.9424 \\
FedAdam v1       & 100\,\% & 6.142 & 42 & 0.9167 & FedProx v2         & 100\,\% & 6.644  & 40 & 0.9246 \\
FedAdam v1       & 20\,\%  & 1.660 & 42 & 0.9325 & FedProx v3         & 20\,\%  & 2.212  & 41 & 0.9389 \\
FedAdam v2       & 20\,\%  & 1.915 & 44 & 0.9108 & FedProx v4         & 20\,\%  & 3.039  & 43 & 0.9358 \\
FedAdam v2       & 100\,\% & 6.268 & 38 & 0.9232 & FedProx v4         & 100\,\% & 8.320  & \textbf{48} & 0.9150 \\
FedAdam v3       & 20\,\%  & 1.950 & 40 & 0.9352 & FedProx v5         & 20\,\%  & 2.115  & 41 & 0.9430 \\
FedAdam v4       & 20\,\%  & 2.032 & 41 & 0.9301 & FedAvg + SMOTE     & 100\,\% & 11.130 & 42 & 0.9205 \\
FedAdam v5       & 20\,\%  & 2.014 & \textbf{49} & 0.9147 & FedAdam v5 + SMOTE & 100\,\% & 12.680 & 37 & \textbf{0.9442} \\
FedAdam v5       & 100\,\% & 10.410 & \textbf{46} & 0.9122 & FedProx v4 + SMOTE & 100\,\% & 6.530  & 42 & 0.9122 \\
FedProx v1       & 20\,\%  & 1.865 & 45 & 0.9241 &                    &         &        &    & \\
\bottomrule
\end{tabular}
\end{table*}

\section{Discussion}\label{sec:disc}

\subsection{Algorithmic mechanisms}
FedAvg produces a general, sample-weighted model with no mechanism to counter heterogeneity. Our adaptive optimizer applies a single Adam-style step per round to the averaged client weights; because the moment estimates reset each round, the server learning rate is the main control. Our proximal optimizer instead constrains each client toward the global model through the client-side penalty or regularization and an aggregation-level damping.

\subsection{Why each optimizer helps}
With $\mu{=}0.5$, the proximal term curbs client-specific overfitting on a single attack class, producing a broadly acceptable global model, so more clients clear the 70\,\% threshold; whereas lower $\mu$ becomes similar to FedAvg. For the adaptive optimizer, a larger server learning rate produces larger steps toward the aggregated direction, resulting in better transferability. These are some of the possible reasons for our observations.

\subsection{The 20\,\%-vs-100\,\% paradox}
We repeatedly observed the 20\,\% stratified subset matching or exceeding the 100\,\% dataset on pair count. We attribute this to potential implicit regularization from smaller samples reducing overfitting to client-specific noise. Stratified 20\,\% subsets thus could be an efficient, representative proxy for hyperparameter search, though final claims should use the full data.

\subsection{The depth-vs-breadth trade-off with SMOTE}
SMOTE's interpolated samples tend to sharpen local minority-class boundaries (boosting per-pair accuracy) but decrease the number of transferable pairs. SMOTE is thus a depth-oriented choice; bootstrapping remains preferable for breadth (more transferable pairs).

\section{Limitations and Future Work}
Our optimizers are simplified adaptations rather than the canonical methods: the adaptive aggregation applies a single Adam step per round over averaged client weights (moments not carried across rounds); the proximal optimizer adds an aggregation-level damping beyond the client penalty. Re-running with persistent server moments, delta-based pseudo-gradients, and multiple seeds could sharpen these comparisons. Following FedAdam and FedProx in their original forms could yield different results. Our evaluation also rests on single-seed runs over 20 communication rounds, which limits statistical strength. Accuracy averaged over transferable pairs can mask per-class failures on an imbalanced dataset, so per-class F1/precision/recall and confusion matrices could be helpful. A next step can be combining these optimizations with the layer-wise Block-Based Smart Aggregation of TabFIDSv2~\cite{jameel2025developing}, and evaluate cross-dataset and cross-backbone transferability (e.g., on UNSW-NB15 and CSE-CIC-IDS 2018) as that work and FetFIDS~\cite{ghosh2025fetfids} began to explore. We welcome extensions and further contributions similar to or based on our work.

\section{Conclusion}
We presented a systematic comparison of adaptive and proximal optimizers (inspired by FedAdam and FedProx) against FedAvg for cross-class transferability in federated NIDS, evaluated on the CIC-IDS 2017 dataset. The proximal optimizer with a strong penalty term yields the best cross-client breadth (48 transferable pairs vs.\ 42 for FedAvg), while the adaptive optimizer with an aggressive server learning rate is competitive (46 pairs). SMOTE produces a depth-vs-breadth trade-off, improving per-pair accuracy but reducing the number of transferable pairs, and bootstrapping remains the stronger balancing choice for breadth. These findings indicate that the choice of federated optimizer is meaningful and complements existing work on data preprocessing and model selection.

\balance

\end{document}